\documentclass[aps, twocolumn,amsmath,amssymb, superscriptaddress, nofootinbib]{revtex4}
\usepackage[pdftex]{graphicx}
 \usepackage{amsmath}
\usepackage[T1]{fontenc}
\usepackage{amssymb}
\usepackage{amsfonts}
\usepackage{bm}
 \usepackage{amsmath} 
  \usepackage{flushend}
\usepackage{color}
\usepackage{lipsum}
\usepackage{amsfonts} 
\usepackage{amssymb, mathrsfs}
\usepackage{braket}
\usepackage{graphicx} 
\usepackage{subfigure}
\usepackage{bbm}
\def\beq{\begin{equation}}
\def\eeq{\end{equation}}
\def\bsp{\begin{split}}
\def\esp{\end{split}}
\def\bea{\begin{eqnarray}}
\def\eea{\end{eqnarray}}
\def\ba{\begin{array}}
\def\ea{\end{array}}

\def\dg{\dagger}

\def\l.{\left.}
\def\r.{\right.}

\def\ra{\rangle}
\def\la{\langle}

\def\bo{{{\vec k}}}

\begin{document}
\date{\today}
\title{Photoinduced Floquet topological magnons in Kitaev magnets}
\author{S. A. Owerre}
\affiliation{Perimeter Institute for Theoretical Physics, 31 Caroline St. N., Waterloo, Ontario N2L 2Y5, Canada.}

\author{Paula Mellado}
\affiliation{Perimeter Institute for Theoretical Physics, 31 Caroline St. N., Waterloo, Ontario N2L 2Y5, Canada.}
\affiliation{School of Engineering and Sciences, Adolfo Ib\'a$\tilde{n}$ez University, Santiago 7941169, Chile}

\author{G. Baskaran}
\affiliation{Perimeter Institute for Theoretical Physics, 31 Caroline St. N., Waterloo, Ontario N2L 2Y5, Canada.}
\affiliation{The Institute of Mathematical Sciences, CIT Campus, Chennai 600 113, India}

\begin{abstract}
We study periodically driven pure Kitaev  model and ferromagnetic phase of the  Kitaev-Heisenberg model on the honeycomb lattice by off-resonant linearly and circularly-polarized lights at zero magnetic field. Using a combination of linear spin wave and Floquet theories,  we show that  the effective   time-independent  Hamiltonians in the off-resonant regime map onto the corresponding anisotropic static spin model, plus a tunable photoinduced magnetic field along the $[111]$ direction, which precipitates Floquet topological magnons and chiral magnon edge modes. They are tunable by the light amplitude and polarization.  Similarly, we show that the thermal Hall effect induced by the Berry curvature of the Floquet topological magnons can also be tuned by the laser field.  Our results pave the way for ultrafast manipulation of topological magnons in irradiated Kitaev magnets, and could play a pivotal role in the investigation of ultrafast magnon spin current generation in Kitaev materials. 
\end{abstract}

\maketitle

\textbf{Introduction.--}
Topological band theory of solid-state materials has dominated many aspects of  condensed-matter physics over the past decade \cite{top3, top4}.  The original concept of topological band theory is rooted in insulating electronic systems possessing a nontrivial gap in their energy band structures. They are characterized by the appearance of gapless chiral edge electron modes traversing the bulk gap, which are topologically protected by the Chern number or the $\mathbb{Z}_2$ index of the bulk bands \cite{top3, top4}. 

Generally, the concept of topological band structure  is independent of the statistical nature of the quasiparticle excitations and therefore is not restricted  to insulating electronic systems. Recently, there has been a tremendous interest in the topological properties of spin excitations in insulating quantum magnets. In fact,  bosonic topological spin excitations (magnons and triplons) have been  studied in many different  insulating quantum magnets \cite{rshin,Zhang,th6,owerre,chern,romh,rchi, cr, mcC, Kitaeva, Kitaevb,flu}, and the appearance of chiral edge modes and bulk Chern number have been demonstrated \cite{Zhang,th6,owerre}. Recently, bosonic topological spin excitations mimicking electronic topological insulators  have been experimentally observed in kagome ferromagnet Cu(1,3-bdc) \cite{rchi}, dimerized quantum magnet SrCu$_2$(BO$_3$)$_2$ \cite{mcC}, and  honeycomb ferromagnet CrI$_3$  \cite{cr}. 

 The Mott-insulating honeycomb Kitaev magnets are currently of great interest \cite{Kitaev1, Kitaev2a, Kitaev2, Kitaev3, Kitaev4, Kitaev4a, Kitaev4b, Kitaev5, kit1,kit2,kit3,kit4,kit5,kit6,kit7,kit8,kit9,kit10, zyou}. Candidate Kitaev materials include  Na$_2$IrO$_3$ and $\alpha$-RuCl$_3$  \cite{Kitaev2, Kitaev3, Kitaev2a, Kitaev4, Kitaev5}.  Recently, topologically protected spin waves have been predicted  in the fully-polarized phase of the pure Kitaev model \cite{Kitaeva} and the  Kitaev-Heisenberg model  \cite{Kitaevb} at high magnetic field. In the former, the topological magnons and chiral edge states present in linear spin-wave approximation survive magnon-magnon interactions and therefore are robust \cite{Kitaeva}.   Indeed, the manipulation of topological magnons and magnon spin currents  is essential for their practical applications  in ultrafast magnetic data storage, magnetic switching, and magnon spintronics \cite{magn}. 
 
{The tremendous interest in topological quantum phases of matter has led to different alternative ways for inducing them in quantum materials}. Recently, irradiated solid-state materials have provided an alternative route to extend the search for topological quantum materials in electronic systems \cite{pho1,pho2,pho3,pho4,pho5,pho5a, pho6}. In this formalism,  topologically trivial systems can be periodically driven to  nontrivial topological systems termed  Floquet topological insulators \cite{pho6, pho3}.  They  have an advantage over their static (equilibrium) topological counterpart, in that their intrinsic properties can be manipulated and different topological phases can be achieved. In irradiated insulating quantum magnets with charge-neutral spin excitations \cite{sowe,kar,ely,claas}, the Floquet physics can emerge from the coupling of the electron spin magnetic dipole moment to the laser electric field  through the time-dependent version of the static Aharonov-Casher phase \cite{aha, spin3}, which  acts as a vector potential or gauge field to the spin current \cite{spin1a}.  In this case, the resulting Floquet physics can reshape the underlying Hamiltonian to stabilize magnetic phases  and provides a promising avenue for  inducing and tuning Floquet topological spin excitations \cite{sowe,kar,ely}, with a direct implication  of generating and manipulating ultrafast spin current using terahertz ({\rm THz})  radiation  \cite{ultra}.  Lately, {\rm THz} electric field amplitude exceeding $100~{\rm MV/cm}$ between $10~{\rm THz}$ ($1~{\rm THz} \sim 4~{\rm meV}$) and $72~ {\rm THz}$  has been reported  \cite{sell}. In  this respect, resonant time-domain {\rm THz} spectroscopy has been recently performed in the  candidate Kitaev material  $\alpha$-RuCl$_3$ \cite{lwu}. 

In this paper,  we propose a tunable mechanism to induce and manipulate topological magnons in irradiated Kitaev magnets at zero  magnetic field. We study the pure  Kitaev model \cite{kita} and the ferromagnetic phase of  the  Kitaev-Heisenberg model, which are already  present in the zero magnetic-field classical phase diagram of the Kitaev-Heisenberg model on the honeycomb lattice \cite{Kitaev2}. Using  linear spin wave and Floquet theories, we show that when  the models are periodically driven by  off-resonant linearly- and circularly-polarized lights, they effectively map onto the corresponding static spin model  plus a tunable photoinduced magnetic field along the $[111]$ direction, which is perpendicular to the honeycomb plane. The photoinduced magnetic field precipitates the existence of Floquet topological magnons and chiral edge modes,  in a similar fashion to a homogeneous magnetic field in the undriven systems \cite{Kitaeva, Kitaevb}. However,  the Floquet topological magnons can be tuned by the amplitude and polarization of the laser field. Likewise,  we demonstrate that  the  resulting Floquet thermal Hall conductivity  can be tuned by the laser field.  The photoinduced magnetic field required to induce magnetic order and Floquet topological magnons  in the pure Kitaev model  lies in the interval $0<h(\mathcal E_0,\phi)<2AS$, where  $\mathcal E_0,\phi$ are the amplitude and polarization of the laser field, $A>0$ { is the overall energy scale of the spin exchange interactions} and $S$ is the spin value. Therefore, $h(\mathcal E_0,\phi)$ is much smaller than the high magnetic field $h>4AS$ required to induce topological magnons in the undriven  pure Kitaev model \cite{Kitaeva}. Interestingly,  the Floquet topological magnons in the irradiated Kitaev magnets do not require an explicit time-reversal symmetry breaking term from the  second-order virtual-photon absorption and emission processes \cite{pho4}, which is strictly required in order to induce Floquet topological states in other irradiated quantum systems \cite{pho5a, sowe,kar, pho4, pho1}.

\begin{figure}
\centering
\includegraphics[width=1\linewidth]{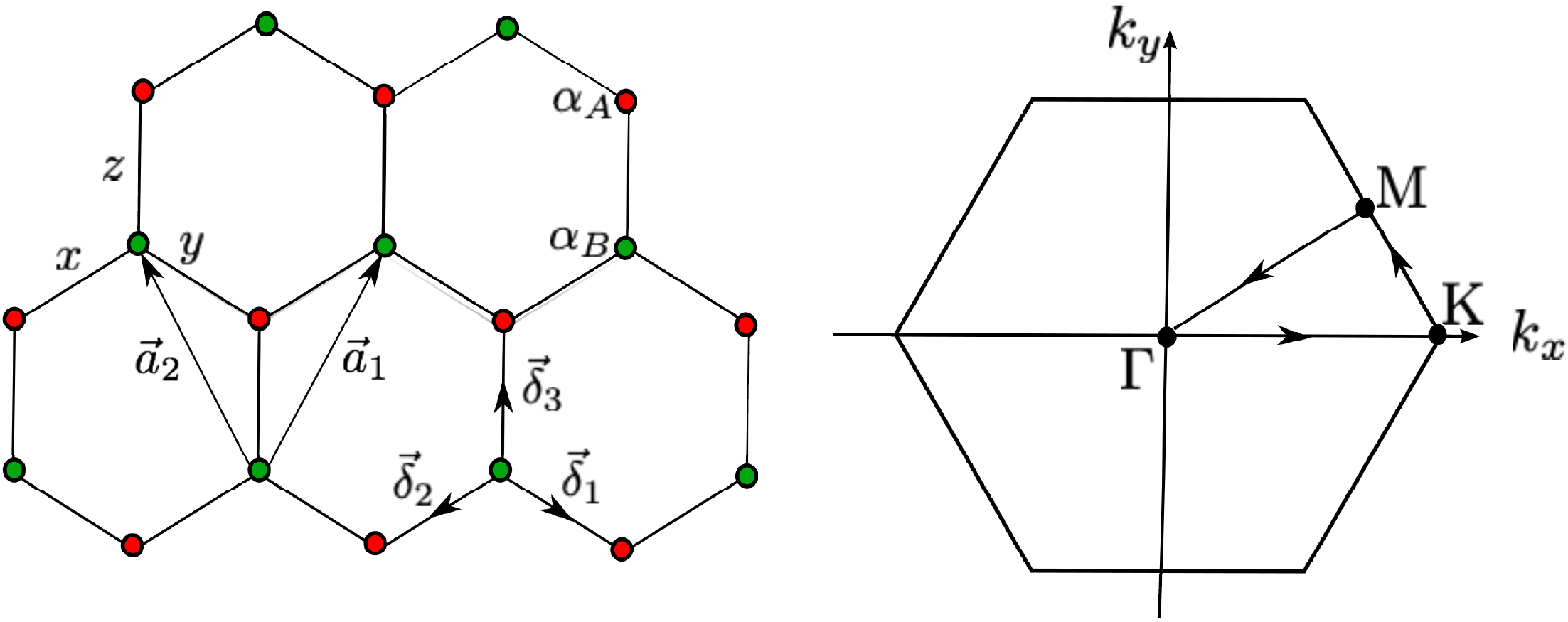}
\caption{Color online.      (a) Honeycomb lattice of the Kitaev model with bond links $\gamma = \lbrace x,y,z\rbrace$ for the Kitaev interaction in Eq.~\eqref{KH}.  The primitive lattice vectors are ${\vec{a}_{1,2}}=(\pm \frac{\sqrt{3}}{2}, \frac{3}{2})a$ and the nearest-neighbour vectors are ${\vec{\delta}_{1}}=(\frac{\sqrt{3}}{2}, -\frac{1}{2})a$, ${\vec{\delta}_{2}}=(-\frac{\sqrt{3}}{2}, -\frac{1}{2})a$, ${\vec{\delta}_{3}}=(0, 1)a$. Here, $\alpha_A$ and $\alpha_B$ denote the two sublattices of the honeycomb lattice. (b) Brillouin zone of the honeycomb lattice with high symmetry points. }
\label{lattice}
\end{figure}

\textbf{Model.--}
We study the Kitaev-Heisenberg  model on the honeycomb lattice with nearest-neighbour interaction. The  spin Hamiltonian reads \cite{Kitaev1, Kitaev2a, Kitaev2, Kitaev3, Kitaev4, Kitaev4a, Kitaev4b, Kitaev5}
\begin{align}
\mathcal H&=   2J_K\sum_{ \la ij\ra \gamma} {S}_{i}^{\gamma}{S}_{j}^{\gamma}+J_H\sum_{\la ij\ra}{\vec S}_{i}\cdot{\vec S}_{j},
\label{KH}
\end{align}
where the first term corresponds to the bond-dependent Kitaev interaction and the second term to the isotropic  Heisenberg interaction.  The bond directions are denoted by $\gamma = \lbrace x,y,z\rbrace$  as shown in Fig.~\eqref{lattice}.  We parameterize the interactions as $J_H=A\cos\vartheta$ and  $J_K= A\sin\vartheta$, where $\vartheta\in [0,2\pi]$ and $A=\sqrt{J_H^2+J_K^2}>0$ is the overall energy scale of the exchange interactions, with $A\sim 8~{\rm meV}$ in some real materials \cite{Kitaev4}. The classical phase diagram of Eq.~\eqref{KH} has been established in the $\vartheta$ space \cite{Kitaev5,Kitaev2}. The zig-zag phase of Eq.~\eqref{KH} is believed to describe the honeycomb magnetic materials  Na$_2$IrO$_3$ and $\alpha$-RuCl$_3$ \cite{Kitaev2, Kitaev4}. Recent studies have shown that the fully-polarized phase of the pure Kitaev model ($\vartheta = \pi/2$) \cite{Kitaeva} and the  Kitaev-Heisenberg model ($\vartheta = 5\pi/4$)  \cite{Kitaevb} at high  magnetic field  possess topological magnon modes.  The purpose of this paper is to periodically drive the magnon topologically trivial phases of Eq.~\eqref{KH} to  Floquet topological magnon insulators for   $\vartheta = \pi/2$ and $\vartheta = 5\pi/4$.

\textbf{Irradiated Kitaev magnets.--} In the presence of  an intense laser field with a dominant time-dependent electric field component $ {\vec {\mathcal E}}(\tau)$, the spin magnetic dipole moment of an electron ${\vec \mu}_S= -g\mu_B\hat{n}$ hopping along the magnetization direction $\hat{n}$ will accumulate a time-dependent Aharonov-Casher phase \cite{ sowe, kar, ely,claas}
\begin{align}
  \Phi_{ij}(\tau) = \mu_m\int_{{\vec r}_i}^{{\vec r}_j}{\vec \Xi}(\tau) \cdot d{\vec \ell},
\end{align}
where $\mu_m = g\mu_B/\hbar c^2$,  $g$ is the spin-g factor, $\mu_B$ is the Bohr magneton,  $\hbar$ is the reduced Plank's constant, and $c$ is the speed of light. Here, ${\vec \Xi}(\tau)= {\vec {\mathcal E}}(\tau) \times \hat{n}$ with $ {\vec {\mathcal E}}(\tau)=-\partial_\tau {\vec {\mathcal A}}(\tau)$, where ${\vec {\mathcal A}}(\tau)$ is the time-dependent vector potential of the applied laser field.

 It is convenient to introduce orthonormal basis vectors $(\hat{l},\hat{m}, \hat{n})$, where $\hat{n}$ points along the cubic $[111]$ direction, perpendicular to the honeycomb plane \cite{chal}. We can now write Eq.~\eqref{KH} in the new basis.  In this new basis, the spin dipole moment of an electron couples to the laser electric field through the Aharonov-Casher phase, in the same way the electron charge couples through the Peierls phase  \cite{pho2,pho4}. Therefore, the terms that contribute to linear spin-wave approximation  can be written as  (see Supplemental material (SM) \cite{sm}) 
 \begin{align}
 \mathcal H(\tau) &= \big(J_H+\frac{2J_K}{3}\big)\sum_{\la ij\ra}\big [S_i^nS_j^n + \frac{1}{2}\lbrace S_i^+S_j^-e^{i\Phi_{ij}(\tau)}+{\rm H.c.}\rbrace\big] \label{rotH}\\&\nonumber +\frac{2J_K}{3}\sum_{\la ij\ra\gamma}\big[\frac{1}{2}\lbrace e^{i\varphi_\gamma}S_i^+S_j^+e^{i\Phi_{ij}(\tau)}+{\rm H.c.}\rbrace\big],
 \end{align}
where $S_j^\pm = S_j^l\pm i S_j^m$ are the usual raising and lowering spin operators, and the angle  $\varphi_\gamma$  comes from the rotation of the bond directions (see SM), with  $\varphi_\gamma = 2\pi/3,4\pi/3,0$ for $x,y,z$  bond directions respectively.  The Aharonov-Casher phase  acts as a vector potential or gauge field to the spin current \cite{spin1a}.  We consider light propagating along the [111] direction ({\it i.e.} perpendicular to the honeycomb plane), given by 
\begin{align}
{\vec \Xi}(\tau)=E_0\big[\sin(\omega \tau),\sin(\omega \tau + \phi), 0\big],
\end{align} 
where $E_0$ is the amplitude of the time-dependent electric field, $\omega$ is the angular frequency of light and $\phi$ is the polarization. Linearly and circularly polarized lights correspond to $\phi =0$  and $\phi=\pi/2$ respectively. {We perform linear spin-wave theory in the polarized phase, which is valid in the large $S$ limit and for low-energy excitations}. This can be done by writing the spin operators in Eq.~\eqref{rotH} in terms of the linearized Holstein-Primakoff  bosons \cite{hp}:  $S_i^n = S-a_i^\dg a_i,~S_i^+ \approx \sqrt{2S}a_i$ for   $i \in \alpha_A$, and $S_j^n = S-b_j^\dg b_j,~S_j^+ \approx \sqrt{2S}b_j$ for   $j \in \alpha_B$. The resulting linear spin-wave bosonic Hamiltonian is time-periodic $\mathcal H_2(\tau+T)=\mathcal H_2(\tau)$, where $T$ is the period of the driving field.

%

 We can now implement the machinery of Floquet theory \cite{floq}, to study the dynamics of irradiated Kitaev magnets.  In the off-resonant limit  $\hbar\omega\gg A$, light simply modifies the band structures \cite{pho4}. The effect of such off-resonant light is captured in a static effective Hamiltonian $\mathcal H_{eff}$ \cite{pho4, pho2}, defined through the evolution Floquet operator $U$ of the system after one period $T= 2\pi/\omega$ as \bea \mathcal H_{eff}= \frac{i}{T}\log(U),\eea where $U=\mathcal{T}\exp\big(-i\int_0^T\mathcal{H}_2(\tau) d\tau\big)$ and $\mathcal T$ is the time-ordering operator.    The effective Hamiltonian can be written as  $\mathcal H_{eff}=\sum_{i\geq 0}\mathcal H_{eff}^{(i)}/(\hbar\omega)^i$.  We work in the off-resonant limit where the photon energy is much larger than the energy scale of the static system, i.e. $\hbar\omega\gg A$. That means we focus on the zero-photon sector \cite{pho2}, $\mathcal H_{eff}^{(0)}=\mathcal{H}_2^0$, where  $\mathcal H_2^n=\frac{1}{T}\int_0^T d\tau e^{-in\omega \tau} \mathcal H_2(\tau)$ are the discrete Fourier components and $n\in \mathbb{Z}$.  Next, we Fourier transform $\mathcal{H}_2^0$ into momentum space and use the basis vector $\big[\psi^{(0)}({ \bo})\big]^\dg=\big(a_{\bo,\alpha_A}^{(0),\dg},b_{\bo,\alpha_B}^{(0),\dg},a_{-\bo,\alpha_A}^{(0)},b_{-\bo,\alpha_B}^{(0)}\big)$. The  effective  time-independent  Hamiltonian is given by 

\begin{align}
 \mathcal H_{eff}^{(0)}({\bo})=S
 \begin{pmatrix}
\mathcal M^{(0)}(\bo)&\mathcal N^{(0)}(\bo)\\ \big[\mathcal N^{(0)}(\bo)\big]^\dg& \big[\mathcal M^{(0)}(-\bo)\big]^T
 \end{pmatrix},
 \label{hamp}
\end{align}

\begin{align}
\mathcal M^{(0)}({\bo})=
 \begin{pmatrix}
 \rho_0^{(0)}& \rho_1^{(0)}(\bo)\\  \rho_1^{(0)}(\bo)^*&  \rho_0^{(0)} 
 \end{pmatrix},
\end{align}

\begin{align}
\mathcal N^{(0)}({\bo})=
 \begin{pmatrix}
  0 & \rho_2^{(0)}(\bo)\\  \rho_2^{(0)}(-\bo)&  0 
 \end{pmatrix},
\end{align}
where
\begin{align}
&\rho_0^{(0)} = -3J_H -2J_K,
\label{kita9}
\end{align}

\begin{align}
\rho_1^{(0)}(\bo) &=  \big(J_H+2J_K/3\big)\big[\mathcal J_0(\mathcal E_0)\nonumber\\&+\mathcal J_0(\mathcal E_+(\phi))e^{i\bo\cdot {\vec a}_1}+\mathcal J_0(\mathcal E_-(\phi))e^{i \bo\cdot {\vec a}_2}\big],
\end{align}

\begin{align}
\rho_2^{(0)}(\bo) &=(2J_K/3)\big[\mathcal J_0(\mathcal E_0)+\mathcal J_0(\mathcal E_+(\phi))e^{i (\bo\cdot {\vec a}_1+ 2\pi/3)}\nonumber\\&+\mathcal J_0(\mathcal E_-(\phi))e^{i (\bo\cdot {\vec a}_2-2\pi/3)}\big],
\end{align}
where  $\mathcal J_\ell(x)$ is the Bessel function of order $\ell \in\mathbb{Z}$, and $\mathcal E_\pm(\phi) = \frac{\mathcal E_0}{2}\sqrt{4\pm 2\sqrt{3}\cos\phi}$.   The dimensionless quantity that characterizes the light intensity is  $\mathcal E_0 = g\mu_BE_0a/\hbar c^2$. The static effective Hamiltonian in Eq.~\eqref{hamp} can be diagonalized  by performing a bosonic Bogoliubov transformation  (see SM).

\begin{figure}
\centering
\includegraphics[width=1\linewidth]{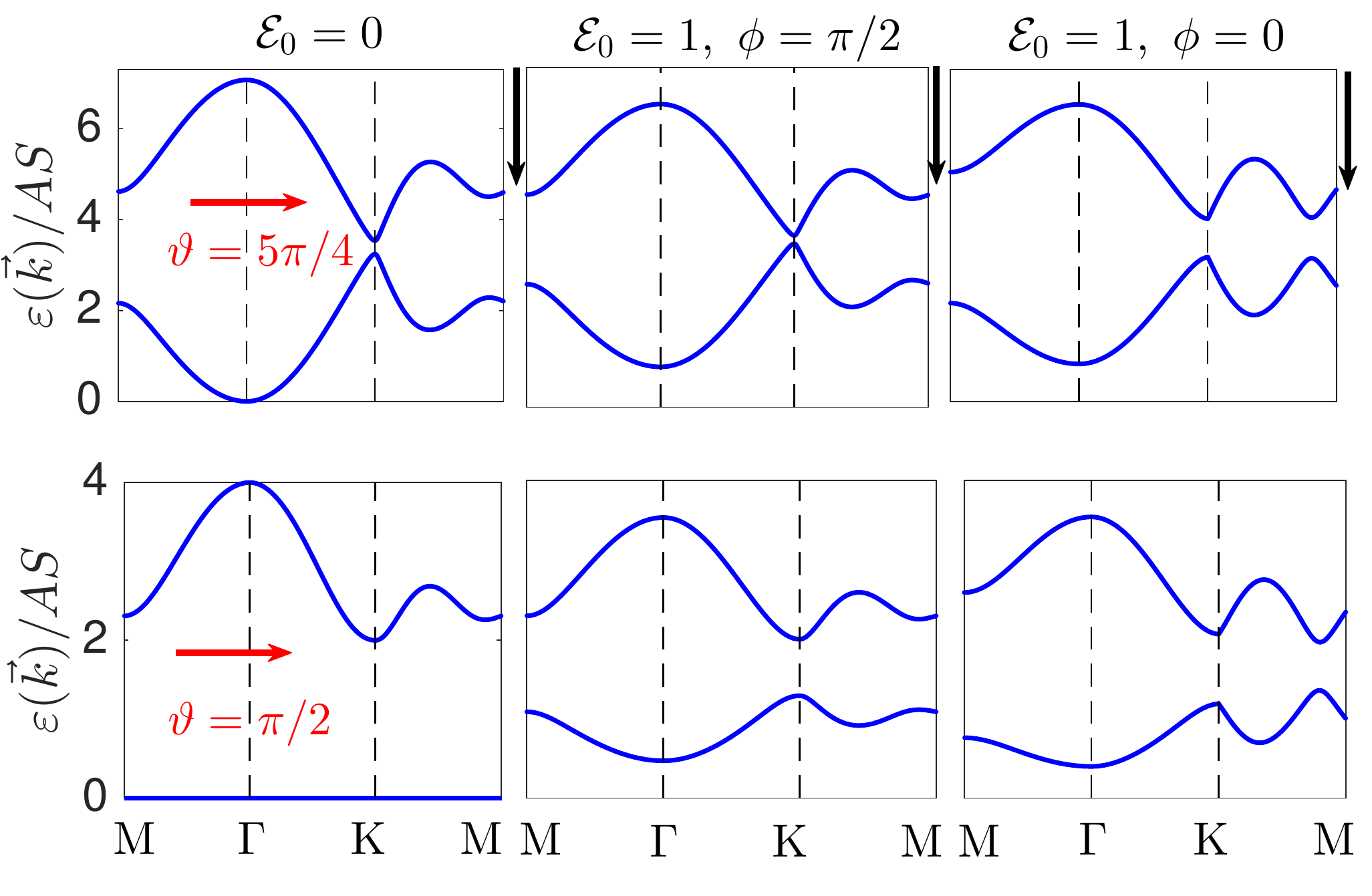}
\caption{Color online. Floquet magnon bands for FM Kitaev-Heisenberg model  $\vartheta = 5\pi/4$ (top panel) and  AFM Kitaev point  $\vartheta = \pi/2$ ( bottom panel).}
\label{band}
\end{figure}

\begin{figure}
\centering
\includegraphics[width=1\linewidth]{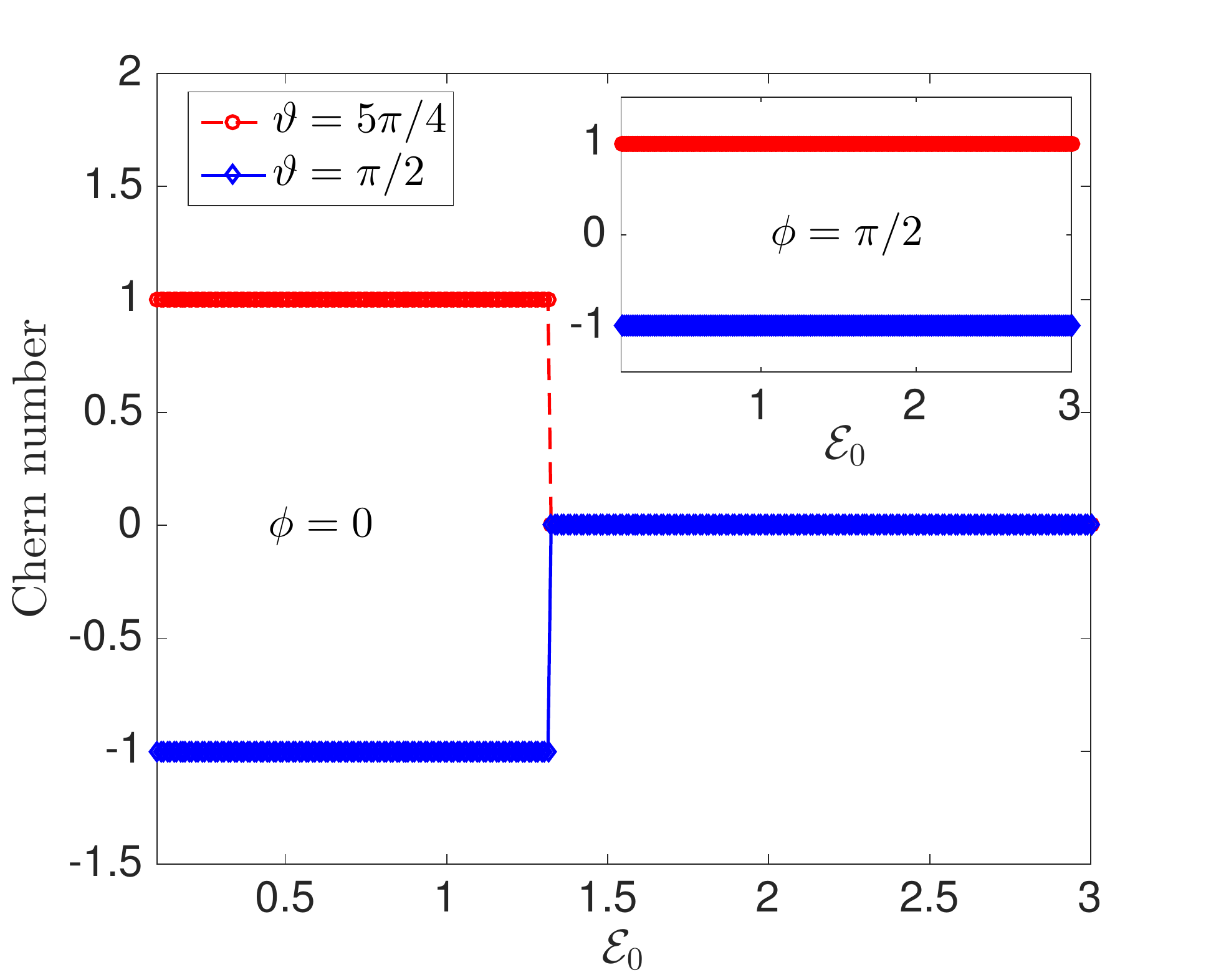}
\caption{Color online. Chern number of the lowest Floquet topological magnon band as a function of $\mathcal E_0$ in  the ferromagnetic Kitaev-Heisenberg model $\vartheta =5\pi/4$ and  at the antiferromagnetic Kitaev  point $\vartheta =\pi/2$ for $\phi=0$. Inset shows the Chern number for  $\phi=\pi/2$.  Note that the  Chern number of the magnon topology is not well-defined at  $\mathcal E_0=0$ (not shown).}
\label{ChernN}
\end{figure}

\textbf{Photoinduced topological magnon bands.--} In Fig.~\eqref{band}, we have shown  the Floquet magnon bands for the ferromagnetic (FM) Kitaev-Heisenberg model (top panel) and at the antiferromagnetic (AFM)  Kitaev point  (bottom panel) for $\mathcal E_0=0$,  $(\mathcal E_0=1,~\phi=\pi/2)$, and $(\mathcal E_0=1,~\phi=0)$. In the FM Kitaev-Heisenberg model $\vartheta = 5\pi/4$ (top panel), the magnon bands for the undriven system at $\mathcal E_0 =0$  are already separated by a finite energy gap at the ${\rm K}$ point. In this case, however,  the magnon topology of the system is  not well-defined and was not discussed in Refs.~\cite{Kitaeva,Kitaevb}. By applying a laser drive, the gap at ${\rm K}$ point does not close, however the system is now driven to a well-defined topological magnon insulator as we will show below.    At the  AFM Kitaev point\footnote{The  antiferromagnetic Kitaev point $\vartheta = \pi/2$  is exactly solvable for spin-$1/2$ in terms of Majorana fermions \cite{kita}  and Jordan-Wigner transformation \cite{yong}.} $\vartheta = \pi/2$ \cite{kita,yong} (bottom panel), the lowest magnon band is a zero energy mode  in the undriven system for $\mathcal E_0 =0$ \cite{bask}. The presence of zero energy mode in the spin wave excitations of frustrated magnets is an artifact of an extensive classical degeneracy, and  points to the onset of a classical spin liquid \cite{cla2}.   As the laser field is applied,  the zero energy mode is lifted for  $\phi=\pi/2$ and $\phi=0$, which implies a photoinduced magnetic order without a high applied magnetic field \cite{Kitaeva}. 
 
 To investigate the magnon topology of the system, we define the Chern number of the Floquet magnon bands as the flux of the Berry curvature threading the entire Brillouin zone (BZ): $\mathcal C_{eff}^\alpha(\mathcal E_0, \phi) = \frac{1}{2\pi}\int_{BZ} d^2 k~\Omega^z_{\alpha}(\bo)$, where $\Omega^z_{\alpha}(\bo)$ is the Berry curvature of the Floquet magnon bands labeled by $\alpha=1,2$ (see SM).  The Chern number has been computed using the discretized BZ method \cite{fuk}.  In the main panel of Fig.~\eqref{ChernN}, we show the evolution of the lowest Floquet Chern number as a function of  $\mathcal E_0$ for $\phi=0$ with $\vartheta= 5\pi/4$ and $\vartheta=\pi/2$.  While the inset shows the Chern number for $\phi=\pi/2$. As we mentioned above, the magnon topology of the system is not well-defined   at equilibrium $\mathcal E_0=0$,  thus we do not consider this case.  For $\phi =0$ and $0<\mathcal E_0\lesssim 1.35$, where $h(\mathcal E_0\sim 1.35, \phi=0)\sim AS$ for $\vartheta=\pi/2$ (see Eq.~\eqref{KH0} below), the Chern number of the lowest band is $\mathcal C_{eff}=+1$   in the  FM Kitaev-Heisenberg model $\vartheta =5\pi/4$ and $\mathcal C_{eff}=-1$ at the AFM Kitaev point $\vartheta =\pi/2$;  but the Chern number is zero for $\mathcal E_0> 1.35$. For $\phi = \pi/2$, the Chern number  is nonzero provided $\mathcal E_0\neq 0$.

\textbf{Effective spin Hamiltonian in real space.--}
 To understand  the origin of the photoinduced topological magnons, we can  map the off-resonant effective static Hamiltonian  in Eq.~\eqref{hamp} back to the real-space spin operators keeping in mind the Holstein-Primakoff  bosons. In the original cubic coordinate system,  the real-space effective static spin Hamiltonian which reproduces  Eq.~\eqref{hamp}  is given by\footnote{Note that  Eq.~\eqref{KH0} is valid in linear spin wave approximation for the magnetically-ordered state considered in this paper. Conversely, the effective static spin Hamiltonian that manifests directly from  Eq.~\eqref{rotH} will be different, because  no specific magnetically-ordered state is assumed in Eq.~\eqref{rotH}.} 

 \begin{align}
 \mathcal H_{eff}^{(0)}&=   \sum_{ \la ij\ra \gamma}J_\gamma(\mathcal E_0,\phi){S}_{i}^{\gamma}{S}_{j}^{\gamma}+\sum_{ \la ij\ra}J_{ij}(\mathcal E_0,\phi) {\vec S}_i\cdot{\vec S}_j \label{KH0}\\&\nonumber + h(\mathcal E_0,\phi)\sum_{i}\big(S_i^x+S_i^y+S_i^z\big),
\end{align} 
which is a renormalized Kitaev-Heisenberg model plus a photoinduced magnetic field along the $[111]$ direction. 
The anisotropic Kitaev interactions are given by $J_z(\mathcal E_0) = 2J_K\mathcal J_0(\mathcal E_0)$, $J_y(\mathcal E_0,\phi) = 2J_K\mathcal J_0(\mathcal E_-(\phi))$, and $J_x(\mathcal E_0,\phi) = 2J_K\mathcal J_0(\mathcal E_+(\phi))$. The Heisenberg interactions are distorted with $J_{ij}(\mathcal E_0)=J_H\mathcal J_0(\mathcal E_0)$ along the vertical ${\vec{\delta}_3}$ bond, $J_{ij}(\mathcal E_0,\phi)=J_H\mathcal J_0(\mathcal E_+(\phi))$ along the diagonal ${\vec{\delta}_1}$ bond, and $J_{ij}(\mathcal E_0,\phi)=J_H\mathcal J_0(\mathcal E_-(\phi))$ along the diagonal ${\vec{\delta}_2}$ bond (see Fig.~\eqref{lattice}).  The photoinduced magnetic field is   given by
\begin{align}
h(\mathcal E_0,\phi) &= (2J_K +3J_H)S\Big[1 -\frac{\mathscr J(\mathcal E_0,\phi)}{3}\Big],
\label{Eq13}
\end{align}
where $\mathscr J(\mathcal E_0,\phi) = \mathcal J_0(\mathcal E_0)+\mathcal J_0(\mathcal E_+(\phi))+\mathcal J_0(\mathcal E_-(\phi))$. Eq.~\eqref{Eq13} stems from the non-renormalized Kitaev-Heisenberg interaction in Eq.~\eqref{kita9}.  Note that Eq.~\eqref{Eq13} vanishes at $\mathcal E_0=0$, hence Eq.~\eqref{KH0} reduces to  Eq.~\eqref{KH}. For $\mathcal E_0\neq 0$, however,  Eq.~\eqref{Eq13}  lies in the interval $0<h(\mathcal E_0,\phi)<(2J_K+3J_H)S$. Thus, at the AFM Kitaev point $\vartheta=\pi/2~(J_H=0)$, the photoinduced magnetic field is $0<h(\mathcal E_0,\phi)<2AS$, which is much smaller than the high homogeneous magnetic field $h>4AS$ required to induce topological magnons in the undriven pure Kitaev model  \cite{Kitaeva}.   On the contrary, at the FM Heisenberg point $\vartheta =\pi~(J_K=0)$,  the effective Hamiltonian \eqref{KH0} is simply a distorted fully-polarized honeycomb ferromagnet, which does not possess any topological magnon modes (see SM). 
  \begin{figure}
\centering
\includegraphics[width=1\linewidth]{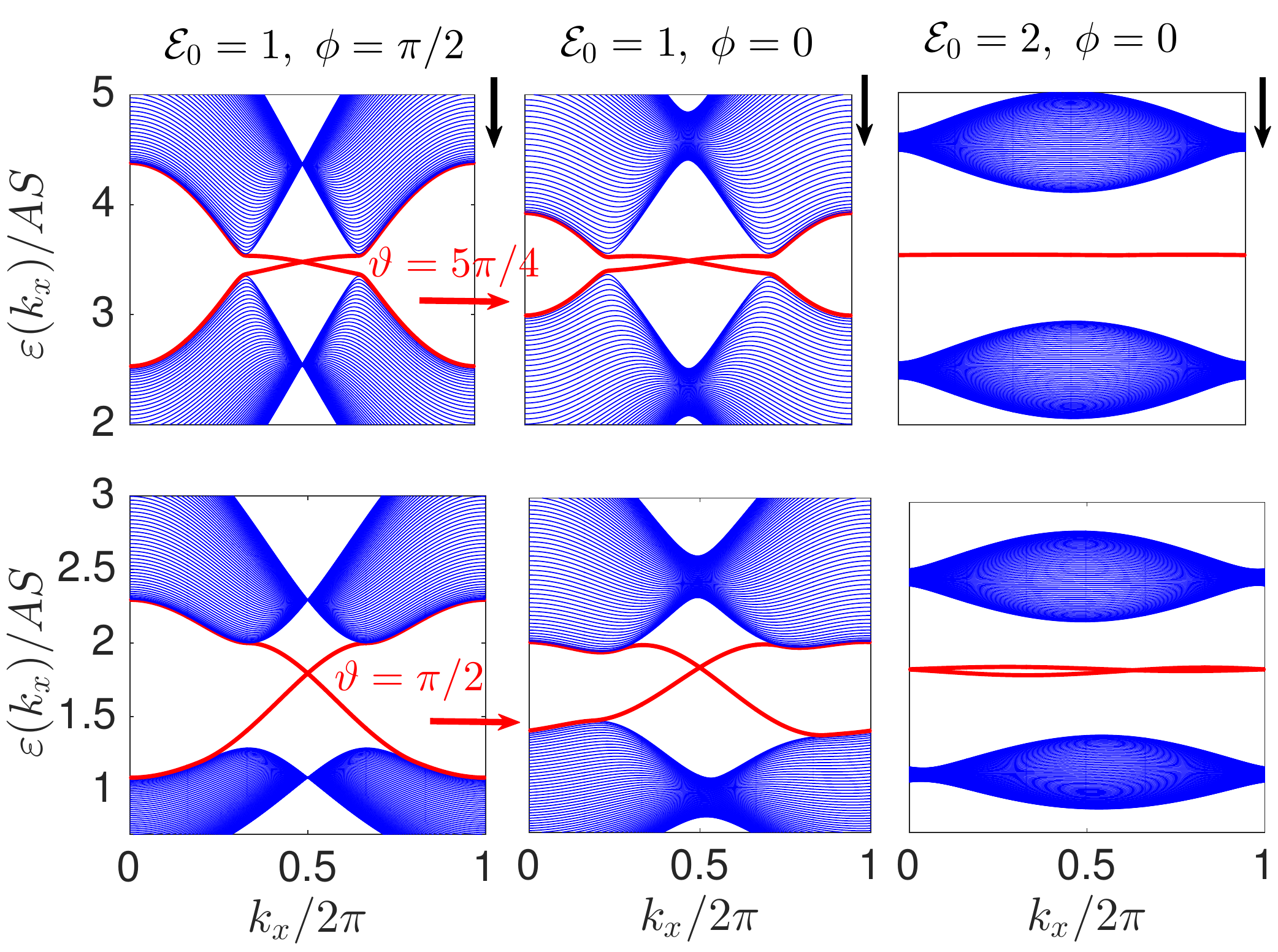}
\caption{Color online. Bulk Floquet magnon bands with tunable zigzag chiral edge states (red curves).  }
\label{Edge}
\end{figure}
\begin{figure}
\centering
\includegraphics[width=1\linewidth]{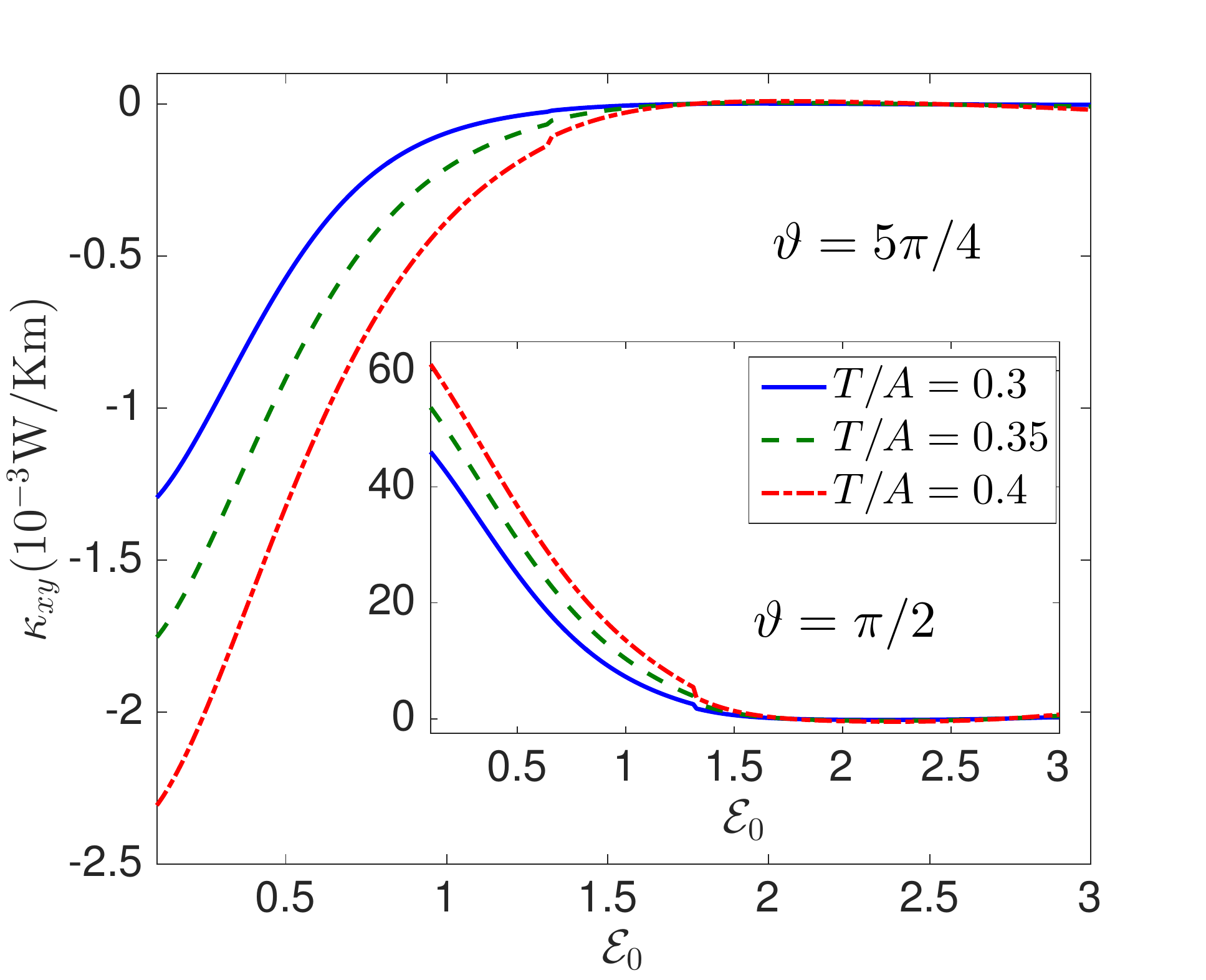}
\caption{Color online. Tunable photoinduced  Floquet thermal Hall conductivity $\kappa_{xy}$ as a function of  $\mathcal E_0$  for $\phi =0$, $S=1/2$  in the Kitaev-Heisenberg model $(\vartheta =5\pi/4)$ and at the AFM Kitaev point $\vartheta=\pi/2$ (inset).}
\label{THE}
\end{figure}

One of the hallmarks of 2D topological systems is the existence of gapless chiral edge modes on the boundary of the system  \cite{top3,top4}.  In insulating topological magnets, the chiral edge modes  can play a pivotal role in spin transport \cite{Zhang}. They are a consequence of the topological properties of the bulk bands. In Fig.~\eqref{Edge}, we show the tunable  zigzag  chiral edge modes (red curves) traversing the bulk gap for $k_x\in \big[\frac{2}{3}\pi,~\frac{4}{3}\pi\big]$  and they cross at the time-reversal invariant momentum $k_x = \pi$ in the topological regime. In the non-topological regime for $\phi=0$ and $\mathcal E_0>1.35$ with $h(\mathcal E_0\sim 1.35, \phi=0)\sim AS$ for $\vartheta=\pi/2$,  the chiral edge modes are completely detached from the bulk bands and they are degenerate along a continuous line, which signifies that the system is topologically trivial as  the Chern number plot in Fig.~\eqref{ChernN} shows.

\textbf{Photoinduced magnon thermal Hall effect.--} The thermal Hall effect  is a consequence of the Berry curvature of topological magnons in magnetically ordered systems \cite{kasa,th1,th2,th5,th7, th4}. In the non-equilibrium Floquet system, we consider the limit where the Bose distribution function of magnon is close to thermal equilibrium. In this limit, the thermal Hall effect  mimics that of equilibrium systems where a longitudinal temperature gradient $-{ \partial}_y { T}$ induces a transverse heat current $ J^q_{x}=-\kappa_{xy}\partial_{y} T$, where $\kappa_{xy}$ is the  thermal Hall conductivity, derived in Ref.~\cite{th5} (see SM \cite{sm}). In Fig.~\eqref{THE}, we show the $\mathcal E_0$-dependence of $\kappa_{xy}$ for $\phi=0$ and  $T/A=0.3,0.35,0.4$, in the FM Kitaev-Heisenberg model $\vartheta =5\pi/4$ and at the AFM Kitaev point $\vartheta =\pi/2$ (inset).  We note that $\kappa_{xy}$ is ill-defined for $\mathcal E_0=0$ at  low temperatures (not shown).  The thermal Hall conductivity is dominated by the Berry curvature of the lowest magnon band at low temperatures and  its sign is consistent with the sign of the Berry curvature (Chern number) of the lowest magnon band.  At low temperature $T/A\ll 1$ and  for $\mathcal E_0> 1.35$, $\kappa_{xy}$ is very small  and approaches zero consistent with the vanishing of the Chern number and the absence of traversing chiral edge modes for $\phi=0$  as shown above.    The low-temperature dependence of  $\kappa_{xy}$ for $\phi=0$ is shown in SM.

\textbf{Conclusion and Outlook.--} We have proposed the existence of  Floquet topological magnon insulators in periodically driven pure Kitaev model and ferromagnetic phase of  the  Kitaev-Heisenberg model at zero  magnetic field.  The main result of our study can be summarized as follows.  In the off-resonant  limit, the Floquet physics stabilizes magnetic order and the  effective time-independent Hamiltonians map onto the corresponding anisotropic static spin model, plus a tunable photoinduced magnetic field along the $[111]$ direction, which facilitates the existence of Floquet topological magnon modes in a similar fashion to a homogenous magnetic field  in the undriven systems \cite{Kitaeva, Kitaevb}. One of the advantages of the current results is that the photoinduced topological magnons and the chiral edge modes can be tuned by varying the amplitude and polarization of the laser field.  Another interesting feature of irradiated Kitaev magnets is that the existence of the Floquet topological magnon insulators does not require the explicit time-reversal symmetry  breaking term from the second-order virtual-photon absorption and emission processes, which is mandatory for  the existence of Floquet topological states in irradiated graphene \cite{pho4, pho1} and irradiated honeycomb ferromagnets  \cite{sowe,kar, ely}.  We also showed that irradiated Kitaev magnets exhibit a tunable photoinduced thermal Hall effect.   A  direct experimental implication  of the current proposal  is that ultrafast magnon spin currents can be generated  in irradiated Kitaev materials using different experimental techniques such as the inverse Faraday effect \cite{ultra} and {\rm THz} spectroscopy \cite{lwu}. This could pave the  way for topological opto-magnonics and  opto-spintronics \cite{magn} using Kitaev materials. 

In future work, we plan to address the effect of magnon-magnon interactions and see how they modify Eq.~\eqref{KH0}. However, it has been shown that the high magnetic-field-induced undriven topological magnons and chiral edge modes present in linear spin-wave approximation remain intact in the presence of magnon-magnon interactions \cite{Kitaeva}. We also plan to study the non-equilibrium distribution function \cite{deh,gbas}  of magnon in this system. Moreover, it would also be interesting to investigate whether tunable topological magnons can be  photoinduced in the zigzag phase of the Kitaev-Heisenberg model.

 \textbf{Acknowledgements.--} 
  Research at Perimeter Institute is supported by the Government of Canada through Industry Canada and by the Province of Ontario through the Ministry of Research and Innovation. PM acknowledges Fondecyt Grant No 1160239.

\end{document}


\begin{widetext}
\setcounter{figure}{0}
\setcounter{equation}{0}
\renewcommand{\thefigure}{S\arabic{figure}}  
\renewcommand{\theequation}{S\arabic{equation}}
\begin{center}
{\bf\large Supplemental Material: Photoinduced Floquet topological magnons in Kitaev magnets}\vspace{0.3cm} \\
{S. A. Owerre,$^{1,2}$ Paula Mellado,$^{1,2}$ G. Baskaran,$^{1,3}$}\\
$^1${\small\it Perimeter Institute for Theoretical Physics, 31 Caroline St. N., Waterloo, Ontario N2L 2Y5, Canada.} \\
$^2${\small\it School of Engineering and Sciences, Adolfo Ib\'a$\tilde{n}$ez University, Santiago 7941169, Chile} \\
$^3${\small\it The Institute of Mathematical Sciences, CIT Campus, Chennai 600 113, India} \\
{\small(Dated: February 1, 2019)}\vspace{0.5cm}\\
\begin{minipage}[ht]{0.8\textwidth}
{\small\hspace{0.3cm}In this Supplemental Material,   we study the effects of the virtual photon absorption term $\mathcal H_{eff}^{(1)}$, and we drive all the tools we used in the main text.}
\end{minipage}
\end{center}

 \section{Kitaev-Heisenberg model in the rotated coordinate system}
 
 It is convenient to introduce the orthonormal basis vectors $(\hat{l},\hat{m}, \hat{n})$, where $\hat{n}$ points along the cubic $(111)$ direction, which is originated perpendicular to the honeycomb plane. The orthonormal basis vectors are related to the cubic-axes basis $\hat{e}_x,\hat{e}_y,\hat{e}_z$ by the transformation 
 
 \begin{align}
 \begin{pmatrix}
 \hat{e}_x\\\hat{e}_y\\ \hat{e}_z
 \end{pmatrix}
 =\begin{pmatrix}
 \frac{1}{\sqrt{3}}&\frac{1}{\sqrt{6}}&-\frac{1}{\sqrt{2}}\\
 \frac{1}{\sqrt{3}}&\frac{1}{\sqrt{6}}&\frac{1}{\sqrt{2}}\\
 \frac{1}{\sqrt{3}}&-\frac{2}{\sqrt{6}}&0
 \end{pmatrix}
  \begin{pmatrix}
 \hat{n}\\\hat{l}\\ \hat{m}
 \end{pmatrix}.
 \end{align}
The spin operators also transform in the same way. Hence, the Kitaev-Heisenberg Hamiltonian in the new coordinate system can be written as \cite{chal}
 
 \begin{align}
 \mathcal H &= \big(J_H+\frac{2J_K}{3}\big)\sum_{\la ij\ra}{\vec S}_i\cdot{\vec S}_j  +\frac{2J_K}{3}\sum_{\la ij\ra\gamma}\big[c_\gamma \big(S_i^lS_j^l-S_i^mS_j^m\big)-s_\gamma \big(S_i^lS_j^m+S_i^mS_j^l\big)\big] \label{rotH}\\&\nonumber-\frac{2J_K\sqrt{2}}{3}\sum_{\la ij\ra\gamma}\big[c_\gamma \big(S_i^lS_j^n+S_i^nS_j^l\big)+s_\gamma \big(S_i^mS_j^n+S_i^nS_j^m\big)\big],
 \end{align}
where ${\vec S}_i=(S_i^l,S_i^m,S_i^n)$, $c_\gamma=\cos\varphi_\gamma$, $s_\gamma=\sin\varphi_\gamma$, and $\varphi_\gamma = 2\pi/3,4\pi/3,0$ for $x,y, z$ bonds respectively.  We can define the spin raising and lowering operators as $S_j^\pm = S_j^l\pm i S_j^m$.  In the rotated frame, the spins couple to light through the time-dependent Aharonov-Casher phase $(S_i^+S_j^-e^{i\Phi_{ij}(\tau)}+{\rm H.c.})$ and $(S_i^+S_j^+e^{i\Phi_{ij}(\tau)}+{\rm H.c.})$, where $\Phi_{ij}(\tau)$ acts as a time-dependent vector potential or gauge field to the spin current.  Note that the last term in Eq.~\eqref{rotH} does not contribute in the linear spin-wave  approximation:  $S_i^n = S-a_i^\dg a_i,~S_i^+ \approx \sqrt{2S}a_i$ 
for   $i \in \alpha_A$, and $S_j^n = S-b_j^\dg b_j,~S_j^+ \approx \sqrt{2S}b_j$ for   $j \in \alpha_B$.
 
\section{Floquet-Bloch theory}
 The linear spin-wave bosonic Hamiltonian in momentum space is time-periodic and can be expanded as
\begin{align}
\mathcal H_2({\vec k},\tau)=\mathcal H_2({\vec k},\tau+T)=\sum_{n=-\infty}^{\infty} e^{in\omega \tau}\mathcal H_2^{n}({\vec k}),
\end{align}
where $\mathcal H_2^{\pm n}({\vec k})=\frac{1}{T}\int_{0}^T e^{\mp in\omega \tau}\mathcal H_2({\vec k}, \tau) d\tau$  are the Fourier components and $T=2\pi/\omega$ is the period of the laser light. The corresponding  eigenvectors of the time-periodic Hamiltonian can be written as 
 \bea 
 \ket{\psi_{\alpha}({\vec k}, \tau)}=e^{-i \varepsilon_{\alpha}({\vec k}) \tau}\ket{u_{\alpha}({\vec k}, \tau)},
 \eea
where  $\ket{u_{\alpha}({\vec k}, \tau)}=\ket{u_{\alpha}({\vec k}, \tau+T)}=\sum_{n} e^{in\omega \tau}\ket{u_{\alpha}^n({\vec k})}$ is the time-periodic Floquet-Bloch wave function of magnon and $\varepsilon_{\alpha}({\vec k})$ are the magnon quasienergy modes. We define the Floquet operator  as $\mathcal H_F({\vec k},\tau)=\mathcal H_2({\vec k},\tau)-i\partial_\tau$. 
The corresponding eigenvalue equation is of the form 
\begin{align}
\sum_m \big[\mathcal H_2^{n-m}({\vec k}) + m\hbar\omega\delta_{n,m}\big]u_{\alpha}^m({\vec k})= \varepsilon_{\alpha}({\vec k})u_{\alpha}^n({\vec k}),
\end{align}
where $n,m$ are integers.  The Fourier components of the single particle Hamiltonian  are given by

\begin{align}
 \mathcal H_2^{(q)}({\bo})=S
 \begin{pmatrix}
\mathcal M^{(q)}(\bo)& \mathcal N^{(q)}(\bo)\\ \big[ \mathcal N^{(-q)}(-\bo)\big]^*& \big[ \mathcal M^{(-q)}(-\bo)\big]^T
 \end{pmatrix},
 \label{fhamp}
\end{align}

\begin{align}
\mathcal M^{(q)}({\bo})=
 \begin{pmatrix}
 \rho_0^q& \rho_1^q(\bo)\\  \rho_1^{-q}(\bo)^*&  \rho_0^q 
 \end{pmatrix},\quad  \mathcal N^{(q)}({\bo})=
 \begin{pmatrix}
  0 & \rho_2^q(\bo)\\  \rho_2^{-q}(-\bo)&  0 
 \end{pmatrix},
\end{align}
%
where $q=n-m$,
\begin{align}
&\rho_0^q = -(3J_H +2J_K)\delta_{q,0},
\end{align}

\begin{align}
\rho_1^q(\bo) &=  \sum_{\ell=1}^3J_{\ell}^F e^{i \bo\cdot {\vec a}_\ell},
\end{align}

\begin{align}
\rho_2^q(\bo) &=  \sum_{\ell=1}^3J_{K,\ell}^F e^{i \bo\cdot {\vec a}_\ell},
\end{align}
where $J_1^F=\big(J_H+2J_K/3\big)\mathcal J_q(\mathcal E_0)e^{iq\phi}$, $J_{2,3}^F=\big(J_H+2J_K/3\big)\mathcal J_{\pm q}(\mathcal E_\pm)e^{\pm iq\Psi_\pm}$. $J_{K,1}^F=(2J_K\mathcal J_q(\mathcal E_0)/3)e^{iq\phi}$, $J_{K,2,3}^F=\mathcal J_{\pm q}(\mathcal E_\pm)\big(J_H+2J_K/3\big)e^{\pm iq\Psi_\pm}e^{\pm i2\pi/3}$, and  $\mathcal J_n(z)$ is the Bessel function of order $n\in \mathbb{Z}$.  
 \begin{align}
 \Psi_\pm = \arctan\lb \frac{\sin\phi}{\sqrt{3}\pm \cos\phi}\rb,\quad \mathcal E_\pm = \frac{\mathcal E_0}{2}\sqrt{4\pm 2\sqrt{3}\cos\phi}
 \end{align}
The dimensionless quantity that characterizes the light intensity is given as  $\mathcal E_0 = g\mu_BE_0a/\hbar c^2$. 
\begin{figure}
\centering
\includegraphics[width=0.9\linewidth]{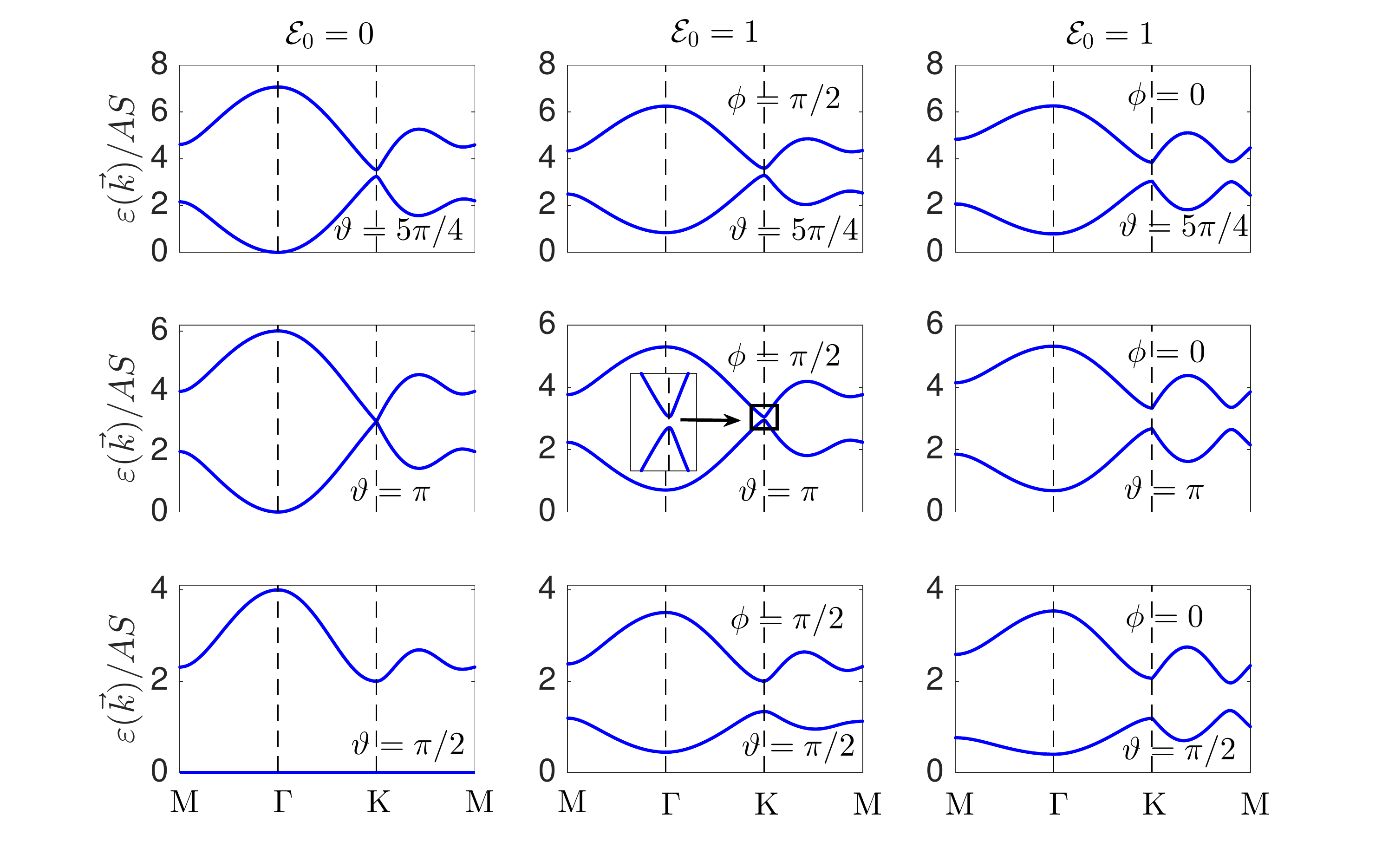}
\caption{Color online. Quasienergy Floquet magnon bands in the FM Kitaev-Heisenberg model  $\vartheta = 5\pi/4$ (top panel), at the FM Heisenberg point $\vartheta=\pi$ (middle panel),  at the  AFM Kitaev point  $\vartheta = \pi/2$ (bottom panel) for $\mathcal E_0=0$, $\mathcal E_0=1$ (vertical panels) and $\phi = \pi/2,0$ (horizontal panels) respectively. The photon energy is $\hbar\omega/AS = 10$.}
\label{SM_band}
\end{figure}

\begin{figure}
\centering
\includegraphics[width=0.75\linewidth]{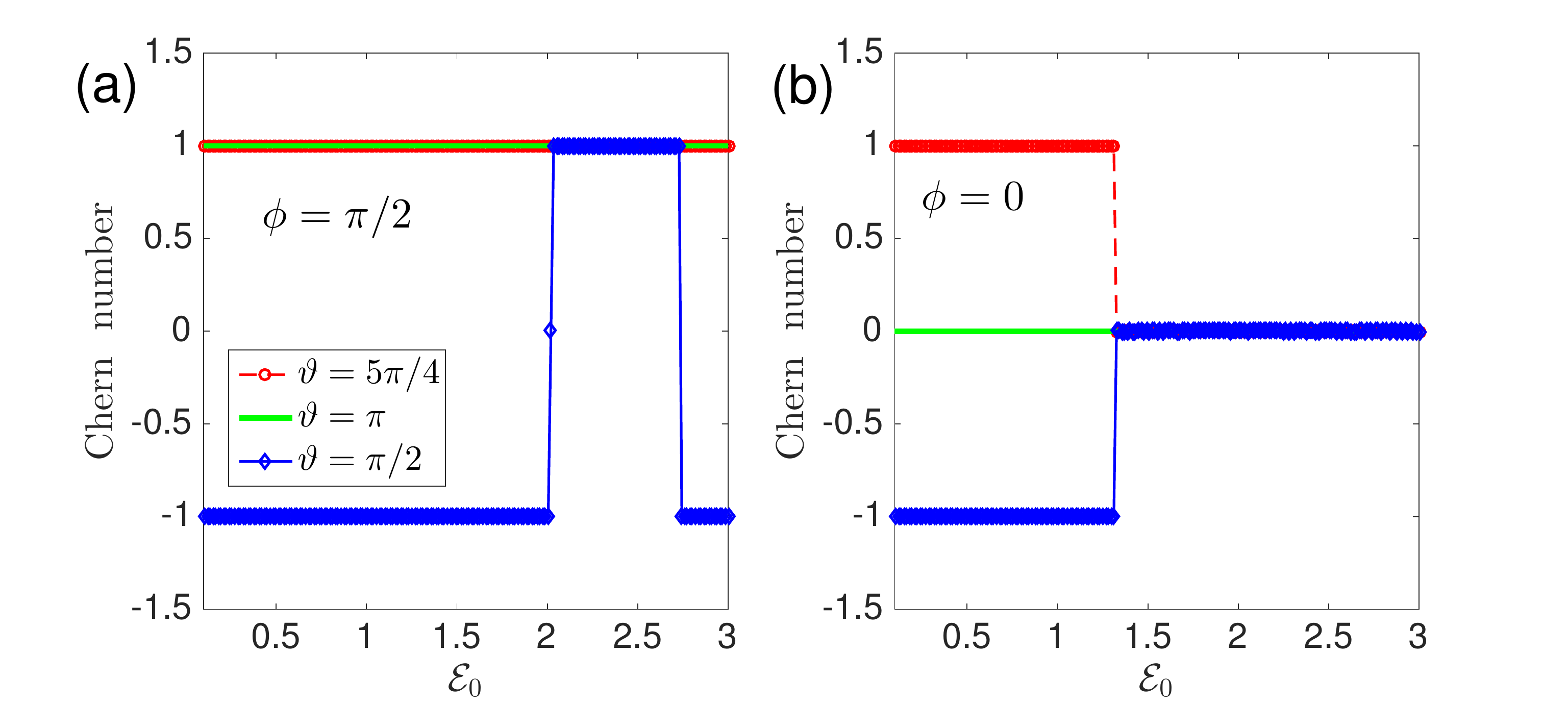}
\caption{Color online. Floquet Chern number of the lowest spin excitation as a function of $\mathcal E_0$ for (a) $\phi=\pi/2$ and  (b) $\phi=0$,  in the FM Kitaev-Heisenberg point ($\vartheta =5\pi/4$),  at the FM Heisenberg point ($\vartheta =\pi$), and  at the AFM Kitaev  point ($\vartheta =\pi/2$). The photon energy is $\hbar\omega/AS = 10$.  The  Chern number of the magnon topology is ill-defined at  $\mathcal E_0=0$ (not shown).}
\label{SM_ChernN}
\end{figure}
 \begin{figure}
\centering
\includegraphics[width=0.75\linewidth]{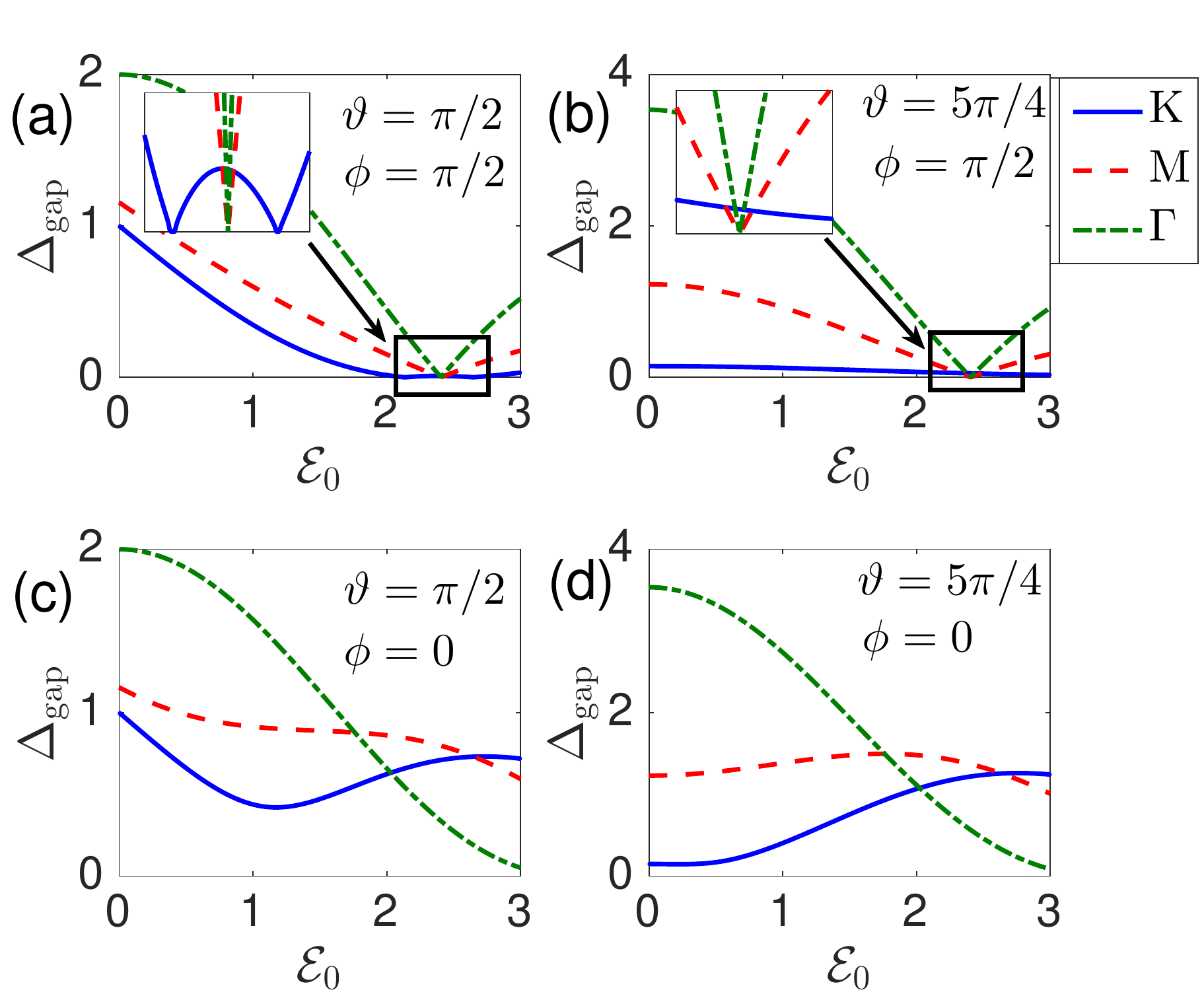}
\caption{Color online. Energy gap (in units of $AS$) as a function of $\mathcal E_0$ at the high symmetry points ${\rm K}$, ${\rm M}$, $\Gamma$. The photon energy is $\hbar\omega/AS = 10$.}
\label{SM_Gap}
\end{figure}

\subsection{Effective time-independent Hamiltonian}

In  the  off-resonant limit $\hbar\omega\gg A$,  the  effective static Hamiltonian can be expanded in power series of $1/\omega$ and the result can be written as

\bea 
\mathcal H_{eff}=\sum_{i\geq 0}\frac{ \mathcal H_{eff}^{(i)}}{(\hbar\omega)^i}.
\eea 
The leading-order term in the zero-photon sector  is  $\mathcal H_{eff}^{(0)}=\mathcal H_2^0$, which is considered in the main text. In this supplemental material, we consider the effect of $\mathcal H_{eff}^{(1)} $ given by
\begin{align}
\mathcal H_{eff}^{(1)} = \sum_{n=1}^{\infty} \frac{ 1}{n}\big[\mathcal H_2^{n}, \mathcal H_2^{-n}\big].
\end{align}
For $n=1$, this term can be understood as the second-order virtual-photon absorption-emission process, where the quasiparticles absorb or emit photon. In terms of spin operators, it is of the form 
\bea
\mathcal H_{eff}^{(1)}\propto  F\Big(J_K,J_H,\mathcal J_1(\mathcal E_0)\Big)\sin\phi\sum_{\la\la jkl \ra\ra,\gamma}\nu_{ij}S_i^{\alpha}S_j^{\gamma}S_k^{\beta},
\label{S15}
\eea  
where the summation runs over the second-nearest neighbour bonds of the honeycomb lattice, and $\nu_{ij}=\pm 1$ for clockwise and anti-clockwise hopping of the quasiparticles respectively.  $  F\Big(J_K,J_H,\mathcal J_1(\mathcal E_0)\Big)$ is a function of interactions and the first-order Bessel function.   Evidently, the results in the main text are unchanged for $\phi=0$, since $\mathcal H_{eff}^{(1)}=0$.  In this supplemental material, we focus on the differences between the antiferromagnetic (AFM) Kitaev point $(\vartheta = \pi/2)$, ferromagnetic (FM)  Kitaev-Heisenberg model $(\vartheta = 5\pi/4)$, and the FM Heisenberg point $(\vartheta = \pi)$.

\subsection{Paraunitary Operator} 

We want to diagonalize the effective Hamiltonian $\mathcal H_{eff}(\bo)=\mathcal H_{eff}^{(0)}(\bo)+\mathcal H_{eff}^{(1)}(\bo)/\hbar\omega$.  The Floquet magnon Hamiltonian $\mathcal{H}_{eff}(\bo)$  can be  diagonalized using the bosonic Bogoliubov transformation \cite{jot}:  $\psi_\bo= \mathcal{P}_\bo Q_\bo$, where $ Q_\bo^\dg=\big(\gamma_{\bo \alpha_A}^{\dg}, \gamma_{\bo \alpha_B}^{\dg}, \gamma_{-\bo \alpha_A}, \gamma_{-\bo \alpha_B}\big)$ are the Bogoliubov quasiparticles, and $\mathcal{P}_\bo$ is a $2N\times 2N$ paraunitary matrix  defined as
\begin{align}
& \mathcal{P}_\bo= \begin{pmatrix}
  u_\bo & v_\bo \\
v_\bo & u_\bo\\  
 \end{pmatrix},
\end{align} 
where $u_\bo$ and $v_\bo$ are  $N\times N$ matrices that satisfy \bea |u_\bo|^2-|v_\bo|^2=I_{N\times N}.\eea  The matrix $\mathcal{P}_\bo$ satisfies the relations,
\begin{align}
 &\mathcal{P}_\bo^\dg  \tau_3 \mathcal{P}_\bo=  \tau_3, \label{eqn1}~
  \mathcal{P}_\bo^\dg \mathcal{H}_{eff}(\bo) \mathcal{P}_\bo=\mathcal{E}(\bo),
\end{align}
where
\begin{align}
\mathcal{E}(\bo) = \begin{pmatrix}\varepsilon_{\alpha}(\bo) & 0\\ 0 & \varepsilon_{\alpha}(-\bo)\end{pmatrix},~  \tau_3=\begin{pmatrix}
I_{N\times N} & 0\\ 0 & -I_{N\times N}
\end{pmatrix},
\end{align}
with quasienergy bands $\varepsilon_{\alpha}(\bo)$ and $N$ is the number of unit cell. This is equivalent to diagonalizing the bosonic Bogoliubov Hamiltonian $\mathcal H_B(\bo)=\tau_3\mathcal H_{eff}(\bo)$. We define  the photoinduced Berry curvature of the bands as 

\begin{align}
\Omega_{\alpha}^{z}(\bo)=-2{\rm Im}\sum_{\beta\neq \alpha}\frac{\big[ \braket{\mathcal{P}_{\bo \alpha}|\hat v_{x}|\mathcal{P}_{\bo \beta}}\braket{\mathcal{P}_{\bo \beta}|\hat v_{y}|\mathcal{P}_{\bo \alpha}}\big]}{\big[ \varepsilon_{\beta}(\bo)-\varepsilon_{\alpha}(\bo)\big]^2},
\label{chern2}
\end{align}
where $\hat v_{x,y}=\partial \mathcal{H}_B(\bo)/\partial k_{x,y}$ defines the photoinduced velocity operators, and $\mathcal{P}_{\bo \alpha}$ are the components of the paraunitary operator. The Berry curvature acts as a fictitious magnetic field in the momentum space, and it acts on magnons in the same way a magnetic field acts on electrons. We can now define the Chern number of the Floquet quasienergy magnon bands as the flux of the Berry curvature threading the entire BZ: \bea \mathcal C_{eff}^\alpha(\mathcal E_0, \phi) = \frac{1}{2\pi}\int_{BZ} d^2 k~\Omega^z_{\alpha}(\bo).\eea The Chern number has been computed using the discretized BZ method \cite{fuk}.

\subsection{Tunable thermal Hall conductivity}
The expression for the tunable thermal Hall conductivity is given by \cite{th5}
\begin{align}
\kappa_{xy}(\mathcal E_0, \phi, T) =- k_BT\int_{BZ} \frac{d\bo}{(2\pi)^2}~ \sum_{\alpha=1}^N c_2\big[ f_\alpha^{B}(\bo)\big]\Omega_{\alpha}^{z}(\bo),
\label{thm}
\end{align}
where $ f_\alpha^{B}(\bo)=\big( e^{\varepsilon_\alpha(\bo)/k_BT}-1\big)^{-1}$ is the Bose-Einstein distribution function of magnon close to thermal equilibrium, $k_B$ the Boltzmann constant which we will  set to unity, $T$ is the temperature  and $ c_2[x]=(1+x)\lb \ln \frac{1+x}{x}\rb^2-(\ln x)^2-2\text{Li}_2(-x)$ is the weight function, with $\text{Li}_2(x)$ being the  dilogarithm. In Fig.~\eqref{SM_THE}, we have shown the temperature dependence of $\kappa_{xy}$ for $\phi=0$ and  $\mathcal E_0=0.5,0.6,0.75$, in the Kitaev-Heisenberg model $\vartheta =5\pi/4$ and at the AFM Kitaev point $\vartheta =\pi/2$ (inset). 
\begin{figure}
\centering
\includegraphics[width=0.6\linewidth]{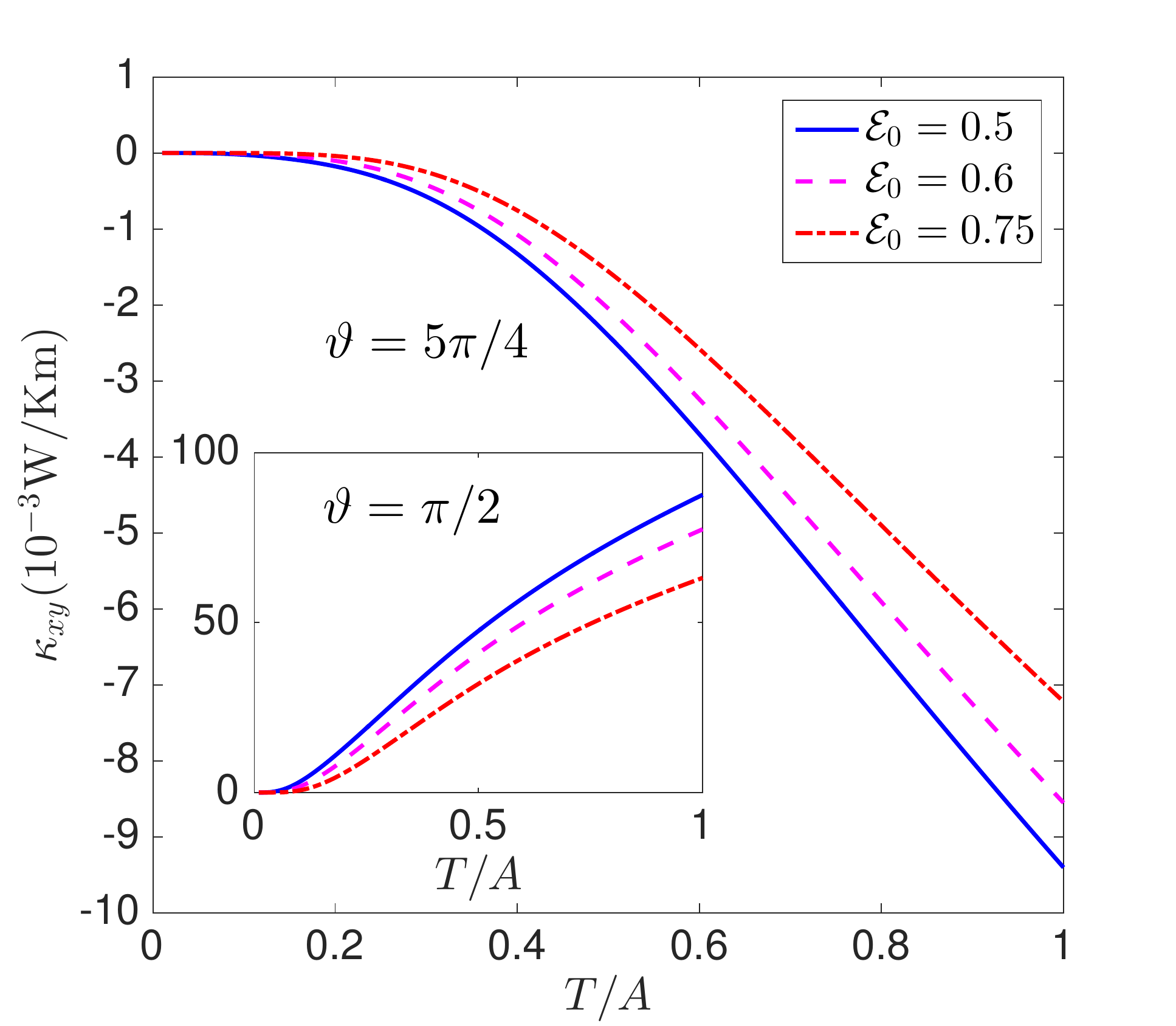}
\caption{Color online. The photoinduced  Floquet thermal Hall conductivity $\kappa_{xy}$ as a function of  $T/A$  at $\phi =0$, $S=1/2$  in the Kitaev-Heisenberg model $(\vartheta =5\pi/4)$ and at the AFM Kitaev point $\vartheta=\pi/2$ (inset). Note that  there is no finite $\kappa_{xy}$ at the  FM Heisenberg point $(\vartheta =\pi)$ for $\phi =0$. Note that $\kappa_{xy}$ is ill-defined at $\mathcal E_0=0$ (not shown)}
\label{SM_THE}

\end{figure}
 \begin{figure}
\centering
\includegraphics[width=0.75\linewidth]{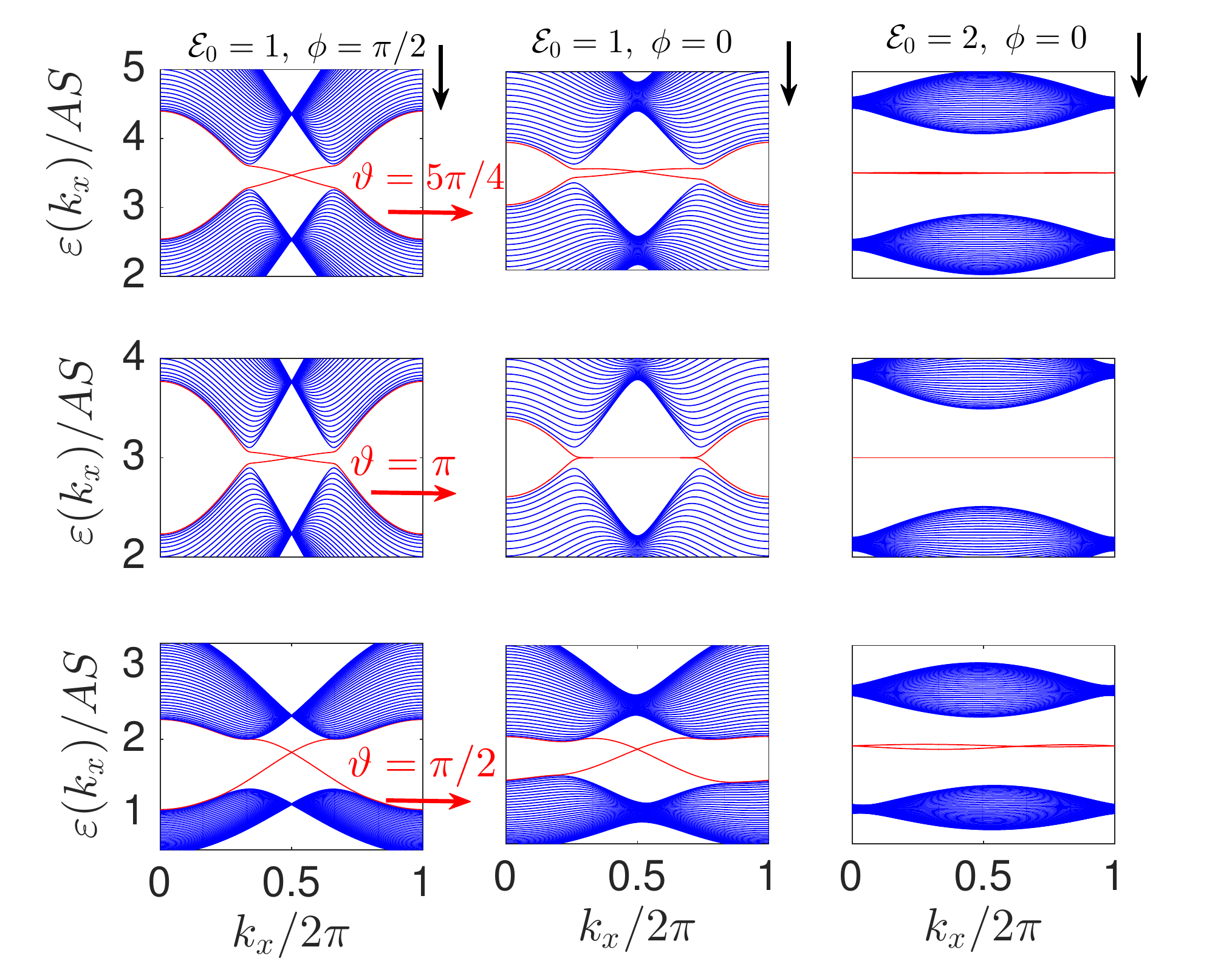}
\caption{Color online. Bulk Floquet magnon bands with zigzag chiral edge states (red curves).  The arrows indicate the directions where the same parameters have been used. The photon energy is $\hbar\omega/AS = 10$.}
\label{SM_Edge}
\end{figure}

\subsection{Photoinduced topological phase transitions}
 In Fig.~\eqref{SM_band}, we have shown  the Floquet quasienergy magnon bands for the Kitaev-Heisenberg model $\vartheta = 5\pi/4$ (top panel),  FM Heisenberg point $\vartheta = \pi$ (middle panel), and  AFM Kitaev point  $\vartheta = \pi/2$ (bottom panel) for $\mathcal E_0=0$,  $(\mathcal E_0=1,~\phi=\pi/2)$, and $(\mathcal E_0=1,~\phi=0)$. In the FM Kitaev-Heisenberg model $\vartheta = 5\pi/4$ (top panel), the magnon bands for the undriven system  are separated by a finite energy gap at the ${\rm K}$ point of the BZ. However, topological properties  at the lowest excitation is not   well-defined \cite{Kitaevb}. The application a laser drive does not close the gap for both linear and circular polarizations, but the system is now driven to a topological phase as we will show below.   At the Heisenberg point $\vartheta = \pi$ (middle panel),  the magnon bands for the undriven system possess Dirac node at the K point. The Dirac node are immediately gap out upon the application of a laser drive due to Eq.~\eqref{S15}  \cite{sowe, kar,ely}. In this case,  the gap opened by linear polarization is topologically trivial as we will show later.  At the  AFM Kitaev point $\vartheta = \pi/2$ (bottom panel), the magnon bands possess zero energy mode \cite{bask, bask1, bask2} for the undriven system. A zero energy mode in the spin wave excitations is well-known in the isotropic  Heisenberg kagome antiferromagnets \cite{cla2, cla3}, and it signifies the existence of an extensive classical degeneracy, {\it i.e.} the onset of a classical spin liquid.   As the laser field is turned on, we can see that the zero energy mode is lifted for both linear and circular polarizations,  which implies a photoinduced magnetic order.  We will now show that the resulting quasienergy magnon bands are topological. 

 In Fig.~\eqref{SM_ChernN}, we have shown the evolution of the lowest Floquet Chern number as a function of  $\mathcal E_0$ for $\phi=\pi/2$ (a) and $\phi=0$ (b) with $\vartheta=\pi/2,\pi,5\pi/4$. We can see that  the Chern number changes sign  as  $\mathcal E_0$ varies at the AFM Kitaev point $\vartheta =\pi/2$. This topological phase transition  is due to $\mathcal H_{eff}^{(1)}(\bo)$.   However, the Kitaev-Heisenberg model $\vartheta =5\pi/4$ and the FM Heisenberg point $\vartheta =\pi$ remain constant as  $\mathcal E_0$ varies. The most distinctive feature of this model is that for linearly polarized light $(\phi =0)$, the Chern number at the FM Heisenberg point $\vartheta =\pi$ vanishes, whereas  the Kitaev-Heisenberg model $\vartheta =5\pi/4$ and the AFM  Kitaev point $\vartheta =\pi/2$ both have non-vanishing Chern number for $0<\mathcal E_0\lesssim 1.35$, but vanishes for $\mathcal E_0> 1.35$.  We note that in Floquet topological systems such as irradiated graphene \cite{pho4} or irradiated Heisenberg ferromagnets \cite{sowe,kar,ely}, linear polarization does not break time-reversal symmetry and thus cannot induce Floquet topological bands. The non-vanishing Chern number in the regime $0<\mathcal E_0\lesssim 1.35$ for linear polarization  is due to the fact that the Kitaev model possesses topological spin excitations at zeroth-order $\mathcal H_ {eff}^{(0)}=\mathcal H^0$ due to photo-induced magnetic field (see main text).

To see why the Chern number changes sign  at the AFM Kitaev point $\vartheta =\pi/2$, we have shown the energy gap in Fig.~\eqref{SM_Gap}. In (a), we can see that the change in the Chern number at AFM Kitaev point $\vartheta=\pi/2$ and  for $\phi=\pi/2$ is accompanied by gap closing and reopening at the ${\rm K}$ point. In  (b) the  Kitaev-Heisenberg model for $\vartheta=5\pi/4$ and $\phi=\pi/2$ does not exhibit any gap closing at the ${\rm K}$ point, hence there is no sign change in the Chern number. In  (c,d), we can see that for $\phi=0$  there is no gap closing for the values of $\vartheta$ we have considered. Hence, the Chern number does not change sign for $\phi=0$.  In  Fig.~\eqref{SM_Edge}, we have shown the bulk bands along a one-dimension zigzag strip. We can see that chiral edge modes (red curves) transverse the bulk gap in the topological regime for $k_x\in \big[\frac{2}{3}\pi,~\frac{4}{3}\pi\big]$, and they cross at the time-reversal invariant momentum $k_x = \pi$. For linear polarization $\phi=0$ and $\mathcal E_0>1.35$, we can see that the chiral edge modes do not cross in the vicinity of the bulk gap, which signifies that the system is topologically trivial as confirmed by the plot of the Chern number in Fig.~\eqref{SM_ChernN}(b).

  \section{Quantum field theory description}
  \label{appenb}
In 2D insulating quantum magnets, topological spin excitations are manifested at the Dirac point located at the ${\rm K}$ point of the BZ. Effectively, the topological properties of the system can be captured at this point in the low-energy limit. As the external electromagnetic field is applied, the system can be  described  by the Dirac-Pauli Lagrangian  \cite{bjo, paul}
\begin{align}
 \mathcal L=\bar\psi\big(-v_0\gamma^0 + i\gamma^\mu\partial_\mu-\frac{\mu_m}{2}\sigma^{\mu\nu} F_{\mu\nu}-m\big)\psi,
 \label{Lag}
\end{align}
where $v_0$ accounts for the finite energy Dirac point and $m$ is the mass gap at the ${\rm K}$ point.  The  two-component wave function is given by $\psi = \big( u_A(\vec x, \tau),u_B(\vec x, \tau) \big)^T$, and $\bar\psi=\psi^\dg\gamma^0$.  Here, $F_{\mu\nu}$  is the The electromagnetic field tensor, and  $\sigma^{\mu\nu}=\frac{i}{2}[\gamma^\mu,\gamma^\nu]=i\gamma^\mu\gamma^\nu,~ (\mu\neq \nu)$ with $\gamma^{\mu}=(\gamma^0,\gamma^i)$. In (2+1) dimensions, one of the representations of gamma matrices is given by
\begin{align}
\gamma^0=\sigma_z,~\gamma^1= i\sigma_y,~\gamma^2= -i\sigma_x,
\end{align}
where $\sigma_i~(i=x,y,z)$ are Pauli matrices. They  satisfy the anticommutation relations $ \lbrace \gamma^\mu,\gamma^\nu\rbrace=2g^{\mu\nu}$, where $g^{\mu\nu}=\text{diag}(1,-1,-1)$ is the Minkowski metric in (2+1) dimensions. The corresponding Hamiltonian is given by
 \begin{align}
 H = \int d^2 x ~\big [ \pi(x)\dot{\psi}(x) - \mathcal L \big ] \equiv \int d^2 x ~ \psi^\dg \mathcal H_D \psi,
 \end{align}
 where $\pi(x) = \partial \mathcal{L}/\partial {\dot\psi (x)}$ 
is the generalized momentum and $x = (\vec x, \tau)$. The Dirac magnon Hamiltonian in the presence of an electromagnetic field is given by
\begin{align}
 \mathcal H_D=v_0{\rm I}_{2\times 2} - i\vec{\alpha}\cdot \vec{\nabla}+\gamma^0\frac{\mu_m}{2}\sigma^{\mu\nu} F_{\mu\nu}+\beta m,
 \label{hamil}
\end{align}
where $\vec{\alpha}=\gamma^0\vec{\gamma}$ and $\beta=\gamma^0$.

Now we consider an electromagnetic field with only a spatially uniform  time-dependent electric field vector $\vec{E}(\tau)$.  In this case,  $\frac{1}{2}\sigma^{\mu\nu} F_{\mu\nu} = i\gamma^0\vec{\gamma} \cdot \vec E(\tau)$. Hence, the corresponding low-energy Dirac Hamiltonian in  momentum space is given by 
\begin{align}
\mathcal H_D({\vec q}, \tau)=v_0{\rm I}_{2\times 2}+\vec{\sigma}_\perp\cdot\big({\vec q}+\mu_m\vec{\Xi}(\tau)\big)+m\sigma_z,
\label{eqn24}
\end{align}
where ${\vec \sigma}_\perp=(\sigma_x,\sigma_y)$ and ${\vec q} = \bo-{\rm K}$ is the momentum deviation from the Dirac point. We can see that the time-periodic electromagnetic field enters the Hamiltonian as a minimal coupling in (2+1) dimensions, in analogy to charged particles. 

\end{widetext}